\documentclass[12pt,a4paper]{article}
\usepackage{cmap}
\usepackage[T2A]{fontenc}
\usepackage[cp1251]{inputenc}
\usepackage[english]{babel}
\usepackage{amsfonts,amsmath,amssymb,amsthm,longtable,hhline}
\usepackage{bm}
\usepackage{cite}
\usepackage{soul}
\usepackage{color,xcolor}
\usepackage[unicode,bookmarks,bookmarksopen,bookmarksopenlevel=2,colorlinks,linkcolor=blue,citecolor=green]{hyperref}
\usepackage{multicol}

\usepackage{tabularx,caption,multirow}
\makeatletter
\renewcommand*\@seccntformat[1]{\csname the#1\endcsname\enspace}
\def\numberline#1{\hb@xt@\@tempdima{#1.\hfil}}
\renewcommand*\l@section[2]{%
	\ifnum \c@tocdepth >\m@ne
	\addpenalty{-\@highpenalty}%
	\vskip 0.2em \@plus\p@
	\setlength\@tempdima{1.5em}%
	\begingroup
	\parindent \z@ \rightskip \@pnumwidth
	\parfillskip -\@pnumwidth
	\leavevmode \bfseries
	\advance\leftskip\@tempdima
	\hskip -\leftskip
	#1\nobreak\mdseries
	\leaders\hbox{$\m@th
		\mkern \@dotsep mu\hbox{.}\mkern \@dotsep
		mu$}\hfill
	\nobreak\hb@xt@\@pnumwidth{\hss #2}\par
	\penalty\@highpenalty
	\endgroup
	\fi}
\makeatother
\makeatletter
\renewcommand{\@seccntformat}[1]{\csname the#1\endcsname. \!}
\renewcommand{\numberline}[1]{\hb@xt@\@tempdima{#1.\hfil}}
\makeatother
\makeatletter
\renewcommand{\@biblabel}[1]{#1.} % Заменяем библиографию с квадратных скобок на точку:
\makeatother
\theoremstyle{remark}

\newcounter{urav}[section]
\newcounter{resh}[urav]

\let\ds=\displaystyle
\let\ts=\textstyle
\let\bl=\bigl \let\br=\bigr
\let\Bl=\Bigl \let\Br=\Bigr
\let\BL=\biggl \let\BR=\biggr
\def\arb{is an arbitrary constant}
\def\arbs{are arbitrary constants}
\def\arbf{is an arbitrary function}
\def\arbfs{are arbitrary functions}
\def\fracskip{\mskip 1mu \relax}
\def\nfrac#1#2{{\fracskip#1\fracskip\over\fracskip#2\fracskip}}
\def\dfrac#1#2{{\ds\nfrac{#1}{#2}}}
\def\tfrac#1#2{{\ts\nfrac{#1}{#2}}}
\let\frac=\nfrac
\def\pd#1#2{\dfrac{\partial#1}{\partial#2}}
\def\pdd#1#2#3{\ifx#2#3\pd{^2#1}{#2^2}\else\pd{^2#1}{#2\partial#3}\fi }

\newcommand{\clh}[1]{\colorbox{yellow}{#1}}%
\newcommand{\clhp}[1]{\colorbox{yellow}{\parbox[t]{\textwidth}{#1}}}%

\let\ds=\displaystyle
\let\ts=\textstyle
\let\bl=\bigl \let\br=\bigr
\let\Bl=\Bigl \let\Br=\Bigr
\let\BL=\biggl \let\BR=\biggr

\newcounter{eexample}
\newcommand{\eexample}[1]{\addtocounter{eexample}{1}
\par \textbf{\textit{Example~\theeexample.}}\, #1 \par}
\newcounter{pproposition}
\newcommand{\pproposition}[1]{\refstepcounter{pproposition}
\par \textbf{Proposition \thepproposition.}\, {\it #1} \par}
\newcounter{rremark}
\newcommand{\rremark}[1]{\refstepcounter{rremark}
\par \textbf{Remark~\therremark.}\, #1 \par}
\newcounter{consequence}
\newcommand{\consequence}[1]{\refstepcounter{consequence}
\par \textbf{Corollary \theconsequence.}\, {\it #1} \par}
\begin{document}

\renewcommand{\contentsname}{\textbf{Contents}\\[-4.5ex]}

\large % увеличение шрифта

\thispagestyle{empty}

%\section*{}
% \hspace{5.00cm}

\begin{center}
\Large\bf Construction of complex solutions to nonlinear partial differential equations using simpler solutions\clh{$^{\star}$}
\end{center}

%\skip

\centerline{Alexander V. Aksenov\,$^{1}$, Andrei D. Polyanin\,$^{2}$}

\bigskip

\centerline{\parbox[t]{0.90\textwidth}{$^{1}$\,\parbox[t]{0.88\textwidth}{Lomonosov Moscow State University, 1 Leninskiye Gory, Main Building,\\ 119234 Moscow, Russia}\\[1.0ex]
$^{2}$\,\parbox[t]{0.88\textwidth}{Ishlinsky Institute for Problems
in Mechanics RAS, 101 Vernadsky \\ Avenue, bldg~1, 119526 Moscow, Russia}}}

\let\thefootnote\relax\footnotetext{
\hskip-20pt\clhp{$^*$ This is a preprint of the article A.V.\! Aksenov, 
A.D.\! Polyanin, Methods for constructing complex solutions of nonlinear PDEs using simpler solutions,  \textit{Mathematics}, 2021, Vol.~9, No.~4, 345; 
doi:\! 10.3390/math9040345.}}

\vspace{5.0ex}

\textbf{Abstract.} The paper describes a number of simple but quite
effective methods for constructing exact solutions of nonlinear partial
differential equations, that involve a relatively small amount of
intermediate calculations. The methods employ two main ideas:
(i)~simple exact solutions can serve to construct more complex
solutions of the equations under consideration and (ii)~exact solutions
of some equations can serve to construct solutions of other, more
complex equations. In particular, we propose a method for constructing
complex solutions from simple solutions using translation and scaling.
We show that in some cases, rather complex solutions can be obtained by
adding one or more terms to simpler solutions. There are situations
where nonlinear superposition allows us to construct a complex
composite solution using similar simple solutions. We also propose a
few methods for constructing complex exact solutions to linear and
nonlinear PDEs by introducing complex-valued parameters into simpler
solutions. The effectiveness of the methods is illustrated by a large
number of specific examples (over 30 in total). These include nonlinear
heat equations, reaction--diffusion equations, wave type equations,
Klein--Gordon type equations, equations of motion through porous media,
hydrodynamic boundary layer equations, equations of motion of a liquid
film, equations of gas dynamics, Navier--Stokes equations, and some
other PDEs. Apart from exact solutions to `ordinary' partial
differential equations, we also describe some exact solutions to more
complex nonlinear delay PDEs. Along with the unknown function at the
current time, $u=u(x,t)$, these equations contain the same function at
a past time, $w=u(x,t-\tau)$, where $\tau>0$ is the delay time.
Furthermore, we look at nonlinear partial functional-differential
equations of the pantograph type, which in addition to the unknown
$u=u(x,t)$, also contain the same functions with dilated or contracted
arguments, $w=u(px,qt)$, where $p$ and $q$ are scaling parameters. We
propose an efficient approach to construct exact solutions to such
functional-differential equations. Some new exact solutions of
nonlinear pantograph-type PDEs are presented. The methods and examples
in the paper are presented according to the principle ``from simple to
complex''. 

\vspace{2.5ex}


\tableofcontents

%\hypertarget{hlink-s:1}{}
\section{Introduction}\label{s:1}

%\hypertarget{hlink-ss:1.1}{}
\subsection{Preliminary Remarks}\label{ss:1.1}

Exact solutions of nonlinear partial differential equations and methods
for their construction are necessary for the development, analysis and
verification of various mathematical models used in natural and
engineering sciences, as well as for testing approximate analytical and
numerical methods. There are several basic methods for finding exact
solutions and constructing reductions of nonlinear partial differential
equations: the method of group analysis of differential equations (the
method of searching for classical
symmetries)~\cite{Ovs,blu1974,ibr1994,olv1989,and1998,BeniaRuggScap},
methods for finding for nonclassical
symmetries~\cite{BlC,lev1989,nuc1992,cher2018}, the direct
Clarkson--Kruskal method~\cite{clar,PolZ2,PolZh,pol2019x}, methods for
generalized separation of variables~\cite{GSv,PolZ2,PolZh}, methods for
functional separation of
variables~\cite{PolZh,pol2019d,pol2020aa,pol2020ab}, the method of
differential constraints~\cite{PolZ2,PolZh,SShYan}, the method of
truncated Painlev\'e expansions~\cite{PolZ2,Kudr2,KM}, and use of
conservation laws to obtain exact
solutions~\cite{Ibr2011,Ibr2012,TracinaBrGT}. The application of these
methods requires considerable special training and, as a rule, is
accompanied by time-consuming analysis and a large volume of analytical
transformations and intermediate calculations.

This paper describes a number of simple, but quite effective, methods
for constructing exact solutions of nonlinear partial differential
equations, which do not require much special training and lead to a
relatively small amount of intermediate calculations. These methods are
based on the following two simple, but very important, ideas:

$\bullet$ simple exact solutions can serve as a basis for constructing
more complex solutions of the equations under consideration,

$\bullet$ exact solutions to some equations can serve as the basis for
constructing solutions to other more complex equations.

The effectiveness of the proposed methods is illustrated by a large
number of specific examples. Nonlinear heat equations,
reaction-diffusion equations, wave type equations, Klein--Gordon type
equations, equations of motion in porous media, hydrodynamic boundary
layer equations, equations of motion of a liquid film, equations of gas
dynamics, Navier--Stokes equations and some other PDEs are considered.
In addition to exact solutions of `ordinary' partial differential
equations, some exact solutions of more complex nonlinear delay PDEs
with constant and variable delay and pantograph-type
functional-differential equations with partial derivatives are also
described.

The methods and examples in the article are presented according to
the principle ``from simple to complex''. For the convenience of a
wide audience with different mathematical backgrounds, the authors
tried to do their best, wherever possible, to avoid special
terminology.

%\hypertarget{hlink-ss:1.2}{}
\subsection{Concept of `Exact Solution' for Nonlinear
PDEs}\label{ss:1.2}

In this article, the term `exact solution' for nonlinear partial
differential equations will be used in cases where the solution is
expressed:

%\begin{itemize}

(i) in terms of elementary functions, functions included in the
equation (this is necessary when the equation contains arbitrary
functions), and indefinite or/and definite integrals;

(ii) through solutions of ordinary differential equations or systems of
such equations.

Combinations of cases (i) and (ii) are also allowed.
In case
(i) the exact solution can be presented in explicit, implicit, or
parametric form.

%\end{itemize}

\rremark{ Exact solutions of nonlinear diffusion and wave type PDEs can
be found, for example, in
\cite{nuc1992,ibr1994,and1998,dor1982,GSv,cher2018,PolZ2,PolZh,pol2019d,%
pol2020aa,pol2020ab,kudr1993,gal1994,doy1998,puc2000,est2002,kap2003,%
van2007,van2012,pol2019km,pol2019kk,grun1992,zhd1994,est2002*,hu2007,%
hua2010a,hua2007,bor2016,pol2019xx,zhu2020}.}

%\hypertarget{hlink-s:2}{}
\section{Construction of Complex Solutions from Simple \\ Solutions by
Shift and Scale Transformations}\label{s:2}

%\hypertarget{hlink-ss:2.1}{}
\subsection{Some Definitions. Simplest Transformations}\label{ss:2.1}

We say that a partial differential equation,
\begin{equation}
F(x,t,u,u_x,u_t, u_{xx},u_{xt},u_{tt},\dots)=0,
\label{a1}
\end{equation}
is invariant with respect to a one-parameter invertible transformation,
\begin{equation}
x=X(\bar{x},\bar{t},\bar{u},a),\quad \
t=T(\bar{x},\bar{t},\bar{u},a),\quad \ u=U(\bar{x},\bar{t},\bar{u},a),
\label{b1}
\end{equation}
if, after substituting expressions~\eqref{b1} in~\eqref{a1}, we obtain
exactly the same equation
\begin{equation*}
F\left(\bar{x},\bar{t},\bar{u},\bar{u}_{\bar{x}},\bar{u}_{\bar{t}},
\bar{u}_{\bar{x}\bar{x}},\bar{u}_{\bar{x}\bar{t}},\bar{u}_{\bar{t}\,\bar{t}},
\dots \right)=0.
\end{equation*}

It is important to note that the free parameter $a$, which can take
values in a certain interval $(a_1,a_2)$, is not included in the
equation~\eqref{a1}.

Transformations that preserve the form of equation~\eqref{a1} transform
a solution of the considered equation into a solution of the same
equation.

A function $I(x, t, u)$ (different from a constant and independent of
$a$) is called an \textit{invariant of transformation}~\eqref{b1} if it
is preserved under this transformation, i.e.
\begin{equation*}
I(x,t,u)=I(\bar{x},\bar{t},\bar{u})
\end{equation*}
for all admissible values of the parameter $a$.

A solution $u=\Phi(x,t)$ of equation~\eqref{a1} is called
\textit{invariant} if, under the transformation~\eqref{b1}, it
transforms into exactly the same solution $\bar u=\Phi(\bar x,\bar t)$.

Further, we will consider only one-parameter transformations of the
form
\begin{align*}
&x=\bar{x}+b_1,\quad t=\bar{t}+b_2,\quad
u=\bar{u}+b_3\quad (\textit{translation});\\
&x=c_1 \bar{x},\quad  t=c_2\bar{t},\quad u=c_3 \bar{u}\quad
(\textit{scaling}),
\end{align*}
and the composition of these transformations. Here $b_n$ and $c_n$
($n=1,\,2,\,3$) are constants depending on the free parameter $a$. Such
transformations will be called the \textit{simplest transformations}.

The following example shows how to determine the invariants of the
simplest transformations and the form of the corresponding invariant
solutions.

\eexample{Consider a transformation consisting of the translation in
$x$ and the scaling in $t$ and $u$:
\begin{equation}
x=\bar{x}-m\ln a, \quad \ t=a\bar{t}, \quad \ u=a^k\bar{u},
\label{eq:100}
\end{equation}
where $k$ and $m$ are some constants. Excluding the parameter $a$, we
find two functionally independent invariants:
\begin{equation}
I_1=x+m\ln t, \quad \ I_2=ut^{-k}.
\label{eq:101}
\end{equation}

If the considered equation is invariant under
transformation~\eqref{eq:100}, then it admits an invariant solution,
which can be represented as $I_2=\varphi(I_1)$ \cite{Ovs,ibr1994} or
$u=t^k\varphi(z)$, where $z=x+m\ln t$. Substituting the resulting
expression into the original equation we arrive at an ordinary
differential equation for the function $\varphi=\varphi(z)$.}

%\hypertarget{hlink-ss:2.2}{}
\subsection{Construction of Complex Solutions from Simpler Solutions. \\
Examples}\label{ss:2.2}

Simple one-term solutions in the form of a product of functions of
different variables are most easily found by the method of separation
of variables (the simplest solutions of this type $u=Ax^\alpha t^\beta$
are easily determined from the equations under consideration by the
method of undefined coefficients). The methods for constructing more
complex solutions based on such solutions are described below.

First we will consider a simple multiplicative separable solution of
the special form
\begin{equation}
u=t^k \varphi_1(x),
\label{d1311}
\end{equation}
where $k$ is some constant and $\varphi_1(x)$ is some function. Such
solutions do not change (are invariant) under the scaling
transformation
%when transforming scaling
\begin{equation}
t=a\bar{t}, \quad \ u=a^k\bar{u}.
\label{d1211}
\end{equation}

Below, in the form of a proposition, we describe a method that allows
us to construct more complex solutions based on one-term solutions of
the form~\eqref{d1311}.

\pproposition{\label{prd3}Let the equation
\begin{equation}
F(t,u,u_x,u_t, u_{xx},u_{xt},u_{tt},\dots)=0,
\label{d1111}
\end{equation}
which does not explicitly depend on the $x$, has a simple
solution of the form~\eqref{d1311} and does not change under the
scaling transformation~\eqref{d1211} (i.e., equation~\eqref{d1111} has
the same property as the original solution~\eqref{d1311}). Then this
equation also has a more complex solution of the form
\begin{equation}
u=t^k\varphi_2(z), \quad \ z=x+m\ln{t},
\label{d14110}
\end{equation}
where $m$  \arb.}

\textbf{Proof.} Equation~\eqref{d1111} does not explicitly depend on
the spatial variable and is invariant under the translation in $x$.
Consider transformation~\eqref{eq:100}, which is a composition of the
translation in $x$ and scaling in $t$ and $u$ (see~\eqref{d1211}).
Transformation~\eqref{eq:100} has two functionally independent
invariants~\eqref{eq:101}. Therefore, the solution invariant under
transformation~\eqref{eq:100} has the form~\eqref{d14110} (see
Example~1).

\rremark{In general, the form of functions $\varphi_1(x)$ and
$\varphi_2(x)$, which are included in the original
solution~\eqref{d1311} and the more complex solution~\eqref{d14110},
respectively, may differ and $\psi_2(z)|_{m=0}\not=\varphi_1(x)$.}

Consider now a simple multiplicative separable solution of the special
form
\begin{equation}
u=x^n\psi_1(t),
\label{d5}
\end{equation}
where $n$ is some constant and $\psi_1(t)$ is some function. Such
solutions do not change (are invariant) under the scaling
transformation
\begin{equation}
x=a\bar{x}, \quad \ u=a^n\bar{u}.
\label{d4}
\end{equation}

A more complex solution than~\eqref{d5} can be obtained by using
Proposition~\ref{prd3}, redefining the constants and variables in
\eqref{d1311}--\eqref{d14110} accordingly. A different, but equivalent
method for constructing a more complex solution is described below,
which is sometimes more convenient to use in practice.

\pproposition{\label{prd2}Let the equation
\begin{equation}
F\left(x,u,u_x,u_t, u_{xx},u_{xt},u_{tt},\dots \right)=0,
\label{d3}
\end{equation}
which does not explicitly depend on $t$, has a simple solution of the
form~\eqref{d5}, and does not change under the scaling
transformation~\eqref{d4} (i.e., equation~\eqref{d3} has the same
property as the original solution~ \eqref{d5}). Then this equation also
has a more complex solution of the form
\begin{equation}
u=e^{-npt}\psi_2(y), \quad \ y=xe^{pt},
\label{d6}
\end{equation}
where $p$  \arb.}

\textbf{Proof.} Equation~\eqref{d3} is invariant under the translation in $t$.
Consider a transformation that is a composition of the
translation in $t$ and the scaling in $x$ and $u$ (see~\eqref{d4}):
\begin{equation}
x=a\bar{x}, \quad \ t=\bar{t}-\frac{1}{p}\ln a, \quad \ u=a^n\bar{u},
\label{d61}
\end{equation}
where $p$ is an arbitrary constant ($p\ne 0$). Transformation~\eqref{d61}
has two functionally independent invariants $I_1=y=xe^{pt}$ and
$I_2=e^{npt}u$. Therefore, a solution invariant with respect to
transformation~\eqref{d61} has the form~\eqref{d6}.

Obviously, equation~\eqref{d3} admits the degenerate
solution~\eqref{d6} with $p=0$.

\rremark{In~\eqref{d6}, the function argument~$y$ is linear in~$x$.
Therefore, solution~\eqref{d6} is easy to differentiate with respect to
$x$. This solution representation should be used for equations which
contain partial derivatives with respect to $x$ of a higher order than
with respect to $t$.}

\eexample{Consider the Boussinesq equation
\begin{equation}
u_t=a(uu_x)_x,
\label{eqg:300}
\end{equation}
which describes the unsteady flow of groundwater in a porous medium
with a free surface~\cite{bou1904}.

Equation~\eqref{eqg:300} has a simple exact solution,
\begin{equation}
u=-\frac{x^2}{6a t},
\label{eqg:301}
\end{equation}
which is simultaneously a solution of two types~\eqref{d1311}
and~\eqref{d5}. Let us consider in order both possibilities of
constructing more complex solutions based on solution~\eqref{eqg:301}.

$1^\circ$. Solution~\eqref{eqg:301} and equation~\eqref{eqg:300} retain
their form under the scaling transformation $t=c\bar{t}$,
$u=\bar{u}/c$. Therefore, by virtue of Proposition~\ref{prd3}
equation~\eqref{eqg:300} admits a more complex exact solution,
\begin{equation*}
u=\frac{\varphi(z)}{t}, \quad z=x+k\ln{t},
\end{equation*}
where the function  $\varphi=\varphi(z)$ satisfies the ordinary
differential equation (hereinafter ODE):
\begin{equation}
k \varphi'_z-\varphi=a(\varphi\varphi_z')_z'.
\label{eq:1005}
\end{equation}

Note that equation~\eqref{eq:1005} for $k=0$ admits a one-parameter
family of solutions in the form of a quadratic polynomial,
\begin{equation*}
\varphi=-\frac{x^2}{6a}+C x-\frac{3aC^2}{2},
\end{equation*}
where $C$  \arb. For $C=0$ this solution coincides with the original
solution~\eqref{eqg:301}.

$2^\circ$. Solution~\eqref{eqg:301} and equation~\eqref{eqg:300} retain
their form also under the scaling transformation $x=c\bar{x}$,
$u=c^2\bar{u}$. Therefore, by virtue of Proposition~\ref{prd2},
equation~\eqref{eqg:300} admits another exact solution
\begin{equation*}
u=e^{-2pt}\psi(y), \quad \ y=xe^{pt},
\end{equation*}
where $p$ \arb \ and the function $\psi=\psi(y)$ is described by the
ODE:
\begin{equation*}
py\psi_y'-2p\psi=a(\psi\psi_y')_y'.
\end{equation*}
}

\eexample{Consider now the Guderley equation
\begin{equation}
u_{xx}=a u_yu_{yy},
\label{eqg:132aa}
\end{equation}
which is used to describe transonic gas flows~\cite{gud1962}.

Equation~\eqref{eqg:132aa} admits a simple exact solution,
\begin{equation}
u=\frac{y^3}{3a x^2},
\label{eqg:132ax}
\end{equation}
which is a special case of two types of solutions~\eqref{d1311}
and~\eqref{d5}. Let us consider in order both possibilities of
constructing more complex solutions based on
solution~\eqref{eqg:132ax}.

$1^\circ$. Solution~\eqref{eqg:132ax} and equation~\eqref{eqg:132aa}
retain their form under the scaling transformation $x=c\bar{x}$,
$u=c^{-2}\bar{u}$. Therefore, by virtue of Proposition~\ref{prd3},
equation~\eqref{eqg:132aa} has a more complex exact solution of the
form
\begin{equation*}
u=x^{-2}\varphi(z), \quad \ z=y+m\ln{x},
\end{equation*}
where the function $\varphi=\varphi(z)$ is described by the second-order
ODE:
\begin{equation*}
m^2 \varphi_{zz}''-5m \varphi_z'+6\varphi=a\varphi_z'\varphi_{zz}''.
\end{equation*}
For $m=0$, this equation admits a one-parameter family of solutions in
the form of a cubic polynomial,
\begin{equation*}
\varphi(z)=\frac{z^3}{3a}+Cz^2+aC^2z+\frac{a^2 C^3}{3},
\end{equation*}
where $C$  \arb. For $C=0$ this solution coincides with the original
solution~\eqref{eqg:132ax}.

$2^\circ$. Solution~\eqref{eqg:132ax} and equation~\eqref{eqg:132aa}
retain their form also under the scaling transformation $y=c\bar{y}$,
$u=c^3\bar{u}$. Therefore, by virtue of Proposition~\ref{prd2}, one can
also obtain another more complex exact solution,
\begin{equation*}
u=e^{-3px}\psi(z), \quad \ z=ye^{px},
\end{equation*}
where $p$ \arb \ and the function $\psi=\psi(z)$ is described by the
ODE:
\begin{equation*}
p^2 z^2\psi_{zz}''-5p^2z\psi_z'+9p^2 \psi=a \psi_z' \psi_{zz}''.
\end{equation*}
}

\eexample{In gas dynamics, there is a nonlinear wave equation,
\begin{equation}
u_{tt}=a(u^b u_x)_x,\quad \ \ b\not=0,
\label{d7}
\end{equation}
which admits a simple exact solution of the form
\begin{equation}
u=a^{-1/b}x^{2/b}t^{-2b}.
\label{eqg:132axy}
\end{equation}
This solution belongs to both classes of solutions~\eqref{d1311}
and~\eqref{d5}. Therefore, based on solution~\eqref{eqg:132axy}, we can
construct two more complex solutions described below.

$1^\circ$. Solution~\eqref{eqg:132axy} and equation~\eqref{d7} are
invariant under the scaling transformation $t=c\bar{t}$,
$u=c^{-2/b}\bar{u}$. By virtue of Proposition~\ref{prd3},
equation~\eqref{d7} has a more complex solution of the form
$$
u=t^{-2/b}\varphi(z),\quad \ z=x+m\ln{t},
$$
where the function $\varphi=\varphi(z)$ is described by the ODE:
\begin{equation*}
m^2\varphi_{zz}''-\frac{m(b+4)}{b}\,\varphi_z'+
\frac{2(b+2)}{b^2}\,\varphi=a(\varphi^b \varphi_z')_z'.
\end{equation*}

$2^\circ$. Solution~\eqref{eqg:132axy} and equation~\eqref{d7} are also
invariant under the scaling transformation $x=c\bar{x}$,
$u=c^{2/b}\bar{u}$. Therefore, by virtue of Proposition~\ref{prd2},
equation~\eqref{d7} admits another solution
$$
u=e^{-2pt/b}\psi(y),\quad \ y=xe^{pt},
$$
where $p$ \arb \ and the function $\psi=\psi(y)$ satisfies the
second-order nonlinear ODE:
\begin{equation*}
p^2y^2\psi_{yy}''+\frac{p^2(b-4)}{b}\,y\psi_y'+
\frac{4p^2}{b^2}\,\psi=a(\psi^b \psi_y')_y'.
\end{equation*}
}

\eexample{The system of boundary layer equations on a flat plate
by introducing a stream function is reduced to one nonlinear
third-order PDE:
\begin{equation}
u_yu_{xy}-u_xu_{yy}=\nu u_{yyy},
\label{d91}
\end{equation}
where $\nu$ is the kinematic viscosity of the fluid~\cite{Schl}.

Equation~\eqref{d91} has a simple solution of the form
\begin{align}
u=\frac{6\nu x}y,
\label{eq:1050}
\end{align}
which generates two more complex solutions.

$1^\circ$. Solution~\eqref{eq:1050} and equation~\eqref{d91} do not
change under the scaling transformation $x=a\bar{x}$, $u=a\bar{u}$.
Therefore, by virtue of Proposition~\ref{prd3}, equation~\eqref{d91}
has a more complex solution of the form
$$
u=x\varphi(z),\quad \ z=y+k\ln{x},
$$
where $k$  \arb \ and the function $\varphi=\varphi(z)$ satisfies the
ODE:
\begin{equation*}
-\varphi\varphi_{zz}''+(\varphi_z')^2=\nu\varphi_{zzz}'''.
\end{equation*}

$2^\circ$. Solution~\eqref{eq:1050} and equation~\eqref{d91} do not
change also under the scaling transformation $y=a\bar{y}$,
$u=\bar{u}/a$. Therefore, by virtue of Proposition~\ref{prd2},
equation~\eqref{d91} admits another solution
$$
u=e^{px}\psi(z), \quad \ z=ye^{px},
$$
where $p$ \arb \ and the function $\psi=\psi(z)$ is described by the
ODE:
\begin{equation*}
p\psi\psi_{zz}''-2p(\psi_z')^2=\nu \psi_{zzz}'''.
\end{equation*}
}

\eexample{Consider a fourth-order nonlinear evolution equation
describing the change in the film thickness of a heavy viscous liquid
moving along a horizontal superhydrophobic surface with a variable
surface tension coefficient,
\begin{equation}
u_t=[(a u^3+bx^{2/3} u^2)(u_x-c(x^2 u_{xx})_x)]_x,
\label{urtol1}
\end{equation}
where $a$, $b$, and $c$ are some constants~\cite{AksSCh1,AksSCh2}.

Equation~\eqref{urtol1} has a simple solution of the form
\begin{align}
u=x^{2/3}f(t),
\label{eq:1080}
\end{align}
where the function $f=f(t)$ is described by the first-order ODE with
separable variables
\begin{equation*}
f_t' =\tfrac{10}{81}(2c+9)f^3(a f+b).
\end{equation*}

Solution~\eqref{eq:1080} and equation~\eqref{urtol1} are invariant
under the scaling transformation $x=k\bar{x}$, $u=k^{2/3}\bar{u}$.
Therefore, by virtue of Proposition~\ref{prd2}, equation~\eqref{urtol1}
also has a more complex solution of the form
$$
u=e^{-2pt/3}\psi(y),\quad \ y=xe^{pt},
$$
where the function $\psi=\psi(y)$ satisfies the fourth-order nonlinear
ODE \cite{AksSCh1,AksSCh2}:
\begin{equation*}
py\psi_y'-\tfrac{2}{3}p\psi=
[(a\psi^3+by^{2/3} \psi^2)(\psi'_y -(cy^2 \psi''_{yy})'_y]'_y\,,
\end{equation*}
where $p$  \arb.
}

\eexample{Consider a $n$th-order nonlinear PDE of the form
\begin{equation}
u_t=u^s F(u_x/u,u_{xx}/u,\dots,u_x^{(n)}/u).
\quad \ s\ne 1.
\label{e:00}
\end{equation}

Equation~\eqref{e:00} has a simple solution,
\begin{align}
u=t^{1/(1-s)}\varphi(x),
\label{eq:2000}
\end{align}
where the function $\varphi=\varphi(x)$ is described by the ODE:
\begin{equation*}
\frac{\varphi}{1-s} = \varphi^s F(\varphi_x'/\varphi,
\varphi_{xx}''/\varphi, \dots,\varphi_x^{(n)}/\varphi).
\end{equation*}

Solution~\eqref{eq:2000} and equation~\eqref{e:00} are invariant under
the scaling transformation $t=a\bar{t}$, $u=a^{1/(1-s)}\bar{u}$.
Therefore, by virtue of Proposition~\ref{prd3}, equation~\eqref{e:00}
also has a more complex solution of the form
$$
u=t^{1/(1-s)}\varphi(z),\quad \ z=x+m\ln{t},
$$
where $m$  \arb \ and the function $\theta=\theta(z)$ satisfies the ODE:
\begin{equation*}
m\varphi_z'+\frac{\varphi}{1-s}=\varphi^s
F(\varphi_z'/\varphi,\varphi_{zz}''/\varphi,\dots,\varphi_z^{(n)}/\varphi).
\end{equation*}
}

\pproposition{\label{prd6} Let the equation
\begin{equation}
F(u,u_x,u_t,u_{xx},u_{xt},u_{tt},\dots)=0,
\label{eq:555}
\end{equation}
which does not explicitly depend on $x$ and $t$ (and
therefore admits the traveling-wave solution~\cite{PolZ2}) does not
change under scaling of the unknown function
\begin{equation}
u=c\bar{u},
\label{d810}
\end{equation}
where $c>0$ \arb. Then this equation admits an exact solution (more
complicated than the traveling-wave solution) of the form
\begin{equation}
u=e^{kt}\varphi(z),\quad \ \,z=px+qt,
\label{d1010}
%u=Ce^{mx}\psi(z),\quad \ z=px+qt,\label{d1010*}
\end{equation}
where $k$, $p$, and $q$  \arbs \ ($pq\not=0$).
}

\textbf{Proof.} Consider a transformation that is a composition of
translations in $x$ and $t$ and scaling of the unknown
function~\eqref{d810}:
\begin{equation}
x=\bar{x}+\frac 1p\ln a, \quad \ t=\bar{t}-\frac{1}{q}\ln a,\quad \
u=a^{-k/q}\bar{u},
\label{d620}
\end{equation}
where $a>0$ \arb \ ($c=a^{-k/q}$), $p$ and $q$ are some constants
($pq\ne 0$). The transformation~\eqref{d620} preserves the form of
equation \eqref{eq:555} and has two functionally independent invariants
$I_1=z=px+qt$ and $I_2=e^{-kt}u$. Therefore, a solution that is
invariant with respect to transformation~\eqref{d620}, can be
represented as~\eqref{d1010}. Solution of the form~\eqref{d1010} is
obtained from the invariant solution by applying the scaling
transformation~\eqref{d810}.

\eexample{Consider the nonlinear heat-type equation
\begin{equation}
u_t=au_{xx}+uf(u_x/u),
\label{d1411}
\end{equation}
where $f=f(\xi)$ is an arbitrary function.

Equation~\eqref{d1411} is invariant under the scaling transformation
\eqref{d810}. Therefore, by virtue of Proposition~\ref{prd6}, this
equation has a solution of the form~\eqref{d1010}, where the function
$\varphi=\varphi(z)$ satisfies the nonlinear ODE:
\begin{equation*}
k\varphi+q\varphi_z'=ap^2\varphi_{zz}'' + \varphi f(p\varphi_z'/\varphi).
\end{equation*}
}

\eexample{Consider a more complex nonlinear PDE of order $n$,
\begin{equation}
u_t=u F(u_x/u,u_{xx}/u,\dots,u_x^{(n)}/u).
\label{e:000}
\end{equation}

Equation~\eqref{e:000} is invariant under the scaling transformation
\eqref{d810}. Therefore, by virtue of Proposition~\ref{prd6}, this
equation has a solution of the form~\eqref{d1010}, where the function
$\varphi=\varphi(z)$ satisfies the nonlinear ODE:
\begin{equation*}
k\varphi+q\varphi_z'=
\varphi F(p\varphi_z'/\varphi,p^2\varphi_{zz}''/\varphi,\dots,
p^n\varphi_z^{(n)}/\varphi),
\end{equation*}
where $F(w_1,w_2,\dots,w_n)$ \arbf.
}

%\hypertarget{hlink-ss:2.3}{}
\subsection{Generalization to Nonlinear Multidimensional
Equations}\label{ss:2.3}

The above Propositions 1--3 allow obvious generalizations to the case of
an arbitrary number of spatial variables.

\eexample{Consider the nonlinear heat equation with $n$ spatial
variables
\begin{equation}
u_t=a\sum_{i=1}^n \pd{}{x_i}\BL(u^k \pd{u}{x_i}\BR),\quad \ \ k\ne 0.
\label{e:04}
\end{equation}

Equation~\eqref{e:04} admits a simple multiplicative separable solution,
\begin{align}
u=t^{-1/k}\varphi(x_1,\dots,x_n),
\label{eq:610}
\end{align}
where the function $\varphi=\varphi(x_1,\dots,x_n)$ satisfies the
stationary equation
\begin{equation*}
-\frac{1}{k}\,\varphi=a\sum_{i=1}^n\pd{}{x_i}\BL(\varphi^k \pd{\varphi}{x_i}\BR).
\end{equation*}

Solution~\eqref{eq:610} and equation~\eqref{e:04} are invariant under
the scaling transformation $t=c\bar{t}$, $u=c^{-k}\bar{u}$. Therefore,
by virtue of Proposition~\ref{prd3}, equation~\eqref{e:04} also has a
more complex solution of the form
$$
u=t^{-1/k}\theta(z_1,\dots z_n), \quad \ z_i=x_i+m_i\ln{t},
$$
where $m_i$ \arbs, and the function $\theta=\theta(z_1,\dots z_n)$
satisfies the stationary equation
\begin{equation*}
-\frac{1}{k}\,\theta+\sum_{i=1}^n m_i\pd{\theta}{z_i}=
a\sum_{i=1}^n\pd{}{z_i}\BL(\theta^k \pd{\theta}{z_i}\BR).
\end{equation*}
}

%\hypertarget{hlink-ss:2.4}{}
\subsection{Generalization to Nonlinear Systems of Coupled
Equations}\label{ss:2.4}

The above Propositions 1--3 can also be used to find exact
solutions of systems of coupled PDEs.

\eexample{Consider the nonlinear system consisting of two coupled
reaction-diffusion equations
\begin{equation}
\begin{aligned}
&u_t=a(u^b u_x)_x+uf(u/v),\\
&v_t=a(v^b v_x)_x+vg(u/v),
\end{aligned}
\label{eqg:017a}
\end{equation}
where $a$ and $b$ are some constants ($b\ne 0$), and $f(z)$ and $g(z)$ are
arbitrary functions.

System of equations~\eqref{eqg:017a} has a simple solution of the form
\begin{equation}
u=x^{2/b}\varphi(t), \quad \ v=x^{2/b}\psi(t),
\label{eq:405}
\end{equation}
where the functions $\varphi=\varphi(t)$ and $\psi=\psi(t)$ are
described by the system of first-order ODEs:
\begin{align*}
&\varphi_t'=\frac{2a (b+2)}{b^2}\varphi^{b+1}+
\varphi f(\varphi/\psi),\\
&\psi_t'=\frac{2a (b+2)}{b^2}\psi^{\mu+1}+
\psi g(\varphi/\psi).
\end{align*}

Solution~\eqref{eq:405} and system of equations~\eqref{eqg:017a} are
invariant under the scaling transformation $x=c\bar{x}$,
$u=c^{2/b}\bar{u}$, $v=c^{2/b}\bar{v}$. Therefore, by virtue of
Proposition~\ref{prd2}, the system of equations~\eqref{eqg:017a} also
has a more complex solution of the form
$$
u =e^{-2mt/b}\Phi(z),\quad \ v = e^{-2mt/b}\Phi(z), \quad \ z=xe^{mt},
$$
where the functions $\Phi=\Phi(z)$ and $\Psi=\Psi(z)$ are described by
the ODE system:
\begin{align*}
&mz\Phi_z'-\frac{2m}{b}\Phi=a(\Phi^b \Phi_z')_z'+\Phi f(\Phi/\Psi),\\
&mz\Psi_z'-\frac{2m}{b}\Psi=a(\Psi^b \Psi_z')_z'+\Psi f(\Phi/\Psi),
\end{align*}
where $m$  \arb.
}

\eexample{Consider another nonlinear system consisting of two coupled
reaction-diffusion equations
\begin{equation}
\begin{aligned}
&u_t=a(u^b u_x)_x+u^{b+1}f(u/v),\\
&v_t=a(v^b v_x)_x+v^{b+1}g(u/v),
\end{aligned}
\label{eqg:017e}
\end{equation}
where $a$ and $b$ are some constants ($b\ne 0$), and $f(z)$ and $g(z)$ are
arbitrary functions.

System of equations~\eqref{eqg:017e}  has a simple solution of the form
\begin{equation}
u=t^{-1/b}\varphi(x), \quad \ v=t^{-1/b}\psi(x),
\label{eq:415}
\end{equation}
where the functions $\varphi=\varphi(x)$ and $\psi=\psi(x)$ are
described by the second-order ODE system
\begin{align*}
&-\frac \varphi b=
a(\varphi^b\varphi_x')_x'+\varphi^{b+1} f(\varphi/\psi),\\
&-\frac \psi b=
a(\psi^b\psi_x')_x'+\psi^{b+1} g(\varphi/\psi).
\end{align*}

Solution~\eqref{eq:415} and system of equations~\eqref{eqg:017e} are
invariant under the scaling transformation $t=c\bar{t}$,
$u=c^{-1/b}\bar{u}$, $v=c^{-1/b}\bar{v}$. By virtue of
Proposition~\ref{prd3}, the system of equations~\eqref{eqg:017e} also
has a more complex solution of the form
$$
u =t^{-1/b}\Phi(z),\quad \ v = t^{-1/b}\Psi(z), \quad \ z=x+m\ln{t},
$$
where $m$  \arb, and the functions $\Phi=\Phi(z)$ and $\Psi=\Psi(z)$
satisfy the system ODE:
\begin{align*}
&-\frac{\Phi}{b}+m\Phi_z'=a(\Phi^b\Phi_z')_z'+\Phi^{b+1} f(\Phi/\Psi),\\
&-\frac{\Psi}{b}+m\Psi_z'=a(\Psi^b\Psi_z')_z'+\Psi^{b+1} g(\Phi/\Psi).
\end{align*}
}

%\hypertarget{hlink-s:3}{}
\section{Construction of Complex Solutions by Adding Terms or \\ Combining
Two Solutions}\label{s:3}

%\hypertarget{hlink-ss:3.1}{}
\subsection{Construction of Complex Solutions by Adding Terms \\ to Simpler Solutions}\label{ss:3.1}

In some cases, simple solutions can be generalized by adding one or
more additional terms to them, which leads to more complex solutions
with generalized separation of variables~\cite{GSv,PolZ2,PolZh}. We
demonstrate the possible course of reasoning in such cases using the
examples of the Boussinesq equation~\eqref{eqg:300} and the Guderley
equation~\eqref{eqg:132aa}.

\eexample{As mentioned earlier, the Boussinesq equation~\eqref{eqg:300}
has a solution with a simple separation of variables (quadratic in $x$,
see~\eqref{eqg:301}), which we write as
\begin{equation}
u=\varphi(t)x^2,\quad \varphi(t)=-1/(6at).
\label{eqg:301*}
\end{equation}

Let's try to find a more complex solution in the form of the sum
\begin{equation}
u(x,t)=\varphi(t)x^2+\psi(t)x^k,\quad k\not=2,
\label{eqg:302}
\end{equation}
whose first term coincides with the solution~\eqref{eqg:301*}. The
second term of formula~\eqref{eqg:302} includes the function $\psi(t)$
and the coefficient $k$, which must be found.

Substituting~\eqref{eqg:302} in~\eqref{eqg:300}, after elementary
transformations we get
\begin{equation}
(\varphi'_t-6a\varphi^2)x^2\!+
[\psi'_t-a(k+1)(k+2)\varphi\psi]x^k\!-ak(2k-1)\psi^2x^{2k-2}\!=0.
\label{eqg:303}
\end{equation}

Since this equality must hold identically for any $x$, the functional
coefficients for various powers of $x$ in~\eqref{eqg:303} must be zero.
Thus, there are two possible cases $k=0$ and $k=1/2$ (both correspond
to the vanishing of the coefficient at $x^{2k-2}$), which must be
considered separately.

$1^\circ$. {\it The first case}.  Substituting $k=0$
into~\eqref{eqg:303}, to define the functions $\varphi=\varphi(t)$ and
$\psi=\psi(t)$, we have the system of ODEs:
\begin{align*}
\varphi'_t-6a\varphi^2=0,\quad \
\psi'_t-2a\varphi\psi=0,
%\label{eqg:303}
\end{align*}
the general solution of which is determined by the formulas
\begin{equation}
\varphi(t)=-\frac 1{6a(t+C_1)},\quad \psi(t)=\frac{C_2}{|t+C_1|^{1/3}},
\label{eqg:304}
\end{equation}
where $C_1$ and $C_2$ \arbs.

$2^\circ$. {\it The second case} (the Barenblatt--Zeldovich dipole
solution~\cite{bar1957}). Substituting $k=1/2$ into~\eqref{eqg:303}, we
obtain a system of ODEs for determining the functions
$\varphi=\varphi(t)$ and $\psi=\psi(t)$:
\begin{equation*}
\varphi'_t-6a\varphi^2=0,\quad \
\psi'_t-\tfrac {15}4a\varphi\psi=0.
%\label{eqg:303}
\end{equation*}
The general solution of this system is
\begin{equation}
\varphi(t)=-\frac 1{6a(t+C_1)},\quad \psi(t)=\frac{C_2}{|t+C_1|^{5/8}}.
\label{eqg:305}
\end{equation}

Given the formulas~\eqref{eqg:302}, \eqref{eqg:304}, \eqref{eqg:305},
as a result, we obtain two three-parameter generalized separable
solutions of equation~\eqref{eqg:300}:
\begin{align*}
&u=-\frac 1{6a(t+C_1)}(x+C_3)^2+\frac{C_2}{|t+C_1|^{1/3}},\\
&u=-\frac 1{6a(t+C_1)}(x+C_3)^2+\frac{C_2}{|t+C_1|^{5/8}}(x+C_3)^{1/2},
\end{align*}
where for the sake of greater generality, an arbitrary translation in
$x$ is additionally added.
}

\rremark{The wave type equation with quadratic nonlinearity
\begin{equation*}
u_{tt}=a(uu_x)_x,
\end{equation*}
also admits solutions of the form~\eqref{eqg:302} with $k=0$ and
$k=1/2$.}

\eexample{Let us now return to the Goderley equation~\eqref{eqg:132aa}.
This equation admits the simple exact solution~\eqref{eqg:132ax}, which we
write in the form
$$
u=f(x)y^3,\quad \ f(x)=1/(3a x^2).
$$

We will look for more complex solutions (with generalized separation of
variables) equation~\eqref{eqg:132aa} in the form
\begin{equation}
u(x,y)=\varphi(x)y^k+\psi(x),
\label{eqg:132ac}
\end{equation}
where the functions $\varphi(x)$ and $\psi(x)$ and the constant
$k\not=0$ are determined in the subsequent analysis
(solution~\eqref{eqg:132ax} is a particular case of
solution~\eqref{eqg:132ac} for $k=3$ and $\psi=0$).

It is important to note that binomial solutions of the form
\eqref{eqg:132ac} are quite often encountered in practice and are the
simplest generalized separable solutions of nonlinear PDEs.

Substituting~\eqref{eqg:132ac} in~\eqref{eqg:132aa}, after rearranging
the terms, we come to the relation
\begin{equation}
\varphi''_{xx}y^k-ak^2(k-1)\varphi^2y^{2k-3}+\psi''_{xx}=0,
\label{eqg:132ad}
\end{equation}
which contains the power functions $y^{k}$ and $y^{2k-3}$ and must be
satisfied identically for any $y$.

Consider two cases: $\psi''_{xx}=0$ and $\psi''_{xx}\not=0$.

$1^\circ$. {\it The first case}. When $\psi''_{xx}=0$ we get a binomial
equation with separable variables, which can be satisfied if we set
\begin{equation}
k=3,\quad \varphi''_{xx}-18a\varphi^2=0.
\label{eqg:132ae}
\end{equation}
The general solution of the autonomous ODE~\eqref{eqg:132ae} can be
represented in the implicit form
$$
x=\pm\int(12a\varphi^3+C_1)^{-1/2}d\varphi+C_2.
$$
Moreover, this equation admits a particular solution of the power form
$\varphi=\frac 1{3a}(x+C_1)^{-2}$, which leads to a three-parameter
exact solution of equation~\eqref{eqg:132aa}:
\begin{equation}
u=\frac 1{3a}(x+C_1)^{-2}y^3+C_2x+C_3,
\label{eqg:132af}
\end{equation}
where $C_1$, $C_2$, and $C_3$ \arbs.

$2^\circ$. {\it Second case}. To balance the function
$\psi''_{xx}\not=0$ with second term in equality~\eqref{eqg:132ad}, we
must set $k=3/2$. As a result, we obtain a binomial equation, which can
be satisfied by setting
\begin{equation*}
\varphi''_{xx}=0, \quad \psi''_{xx}=\tfrac98a\varphi^2.
%\label{eqg:132ag}
\end{equation*}
These equations are easily integrated and lead to a four-parameter
exact solution of equation~\eqref{eqg:132aa}:
\begin{equation}
u=(C_1x+C_2)y^{3/2}+\frac{3a}{32C_1^2}(C_1x+C_2)^4+C_3x+C_4,
\label{eqg:132ah}
\end{equation}
where $C_1$, $C_2$, $C_3$, and $C_4$ \arbs.
}

\eexample{Let us return to the hydrodynamic boundary layer
equation~\eqref{d91}. It is easy to verify that this equation admits
the self-similar solution~\cite{pav1961}:
\begin{equation}
u=F(\xi),\quad \ \xi=y/x,
\label{eqg:132ay}
\end{equation}
where the function $F=F(\xi)$ satisfies the third-order ODE:\
$-(F'_z)^2=\nu F'''_{zzz}$.

We look for a more general solution of equation~\eqref{d91} by adding
the function $\varphi(x)$ to \eqref{eqg:132ay}:
\begin{equation*}
u=F(\xi)+\varphi(x),\quad \ \xi=y/x.
\end{equation*}
Simple calculations show that $\varphi(x)=a\ln x$, where $a$ \arb. As a
result, we obtain a non-self-similar solution of the boundary layer
equation ~\eqref{d91} of the form~\cite{PolZ2}:
\begin{equation*}
u=F(\xi)+a\ln x,\quad \ \xi=y/x,
%\label{eqg:132az}
\end{equation*}
where the function $F=F(\xi)$ is described by the third-order ODE: \
$-(F'_z)^2-aF''_{zz}=\nu F'''_{zzz}$. }

%\hypertarget{hlink-ss:3.2}{}
\subsection{Construction of Compound Solutions (Nonlinear Superposition \\ of Solutions)}\label{ss:3.2}

In some cases, two similar but different solutions of the considered
nonlinear PDE can be combined to obtain a more general composite
solution. We demonstrate the possible course of reasoning in such cases
by examples of the Goderley equation and the nonlinear diffusion
equation with a second-order volume reaction.

\eexample{From expressions~\eqref{eqg:132af} and~\eqref{eqg:132ah} it
follows that the Guderley equation~\eqref{eqg:132aa} has two solutions
of the same type $u_1=\varphi y^{3/2}+\nobreak\psi$ and $u_2=\varphi
y^{3}+\psi$, which differ from each other by the exponent $y$. This
circumstance suggests an attempt to construct a more general solution
of equation~\eqref{eqg:132aa}, that includes both terms with different
exponents at once. In other words, we are looking for a
composite solution of the form
\begin{equation}
u(x,y)=\varphi_1(x)y^{3}+\varphi_2(x)y^{3/2}+\psi(x).
\label{eqg:312ba}
\end{equation}
Substitute it in equation~\eqref{eqg:132aa}. After combining the functional factors
for power-functions $y^{3n/2}$ ($n=0,\,1,\,2$), we get
\begin{equation*}
(\varphi_1''-18a\varphi_1^2)y^3+
(\varphi_2''-\tfrac {45}4a\varphi_1\varphi_2)y^{3/2}+\psi''-\tfrac 98a\varphi_2^2=0.
%\label{eqg:312bb}
\end{equation*}
For this equality to hold for any $y$, it is necessary to equate the
functional factors of $y^{3n/2}$ to zero. As a result, we arrive
at the system of ODEs:
\begin{equation}
\begin{aligned}
&\varphi_1''-18a\varphi_1^2=0,\\
&\varphi_2''-\tfrac {45}4a\varphi_1\varphi_2=0,\\
&\psi''-\tfrac 98a\varphi_2^2=0.
\end{aligned}
\label{eqg:312bbb}
\end{equation}

Thus, it is constructively proved that equation~\eqref{eqg:132aa}
admits a solution of the form~\eqref{eqg:312ba} (this solution was
obtained in~\cite{tit1988}).

It can be shown that the system~\eqref{eqg:312bbb} admits the exact
solution
\begin{align*}
&\varphi_1=\frac 1{3a}(x+C_1)^{-2},\\
&\varphi_2=C_2(x+C_1)^{5/2}+C_3(x+C_1)^{-3/2},\\
&\psi =\frac{3a}{112}C_2^2(x+C_1)^7+\frac38aC_2C_3(x+C_1)^3+
\frac 9{16}aC_3^2(x+C_1)^{-1}+C_4x+C_5.
\end{align*}
}

\eexample{Let us now consider a nonlinear diffusion equation with
the second-order volume reaction
\begin{equation}
u_t=a(uu_x)-bu^2.
\label{eqg:306}
\end{equation}
The procedure for constructing a composite solution of this equation
will be carried out in two stages: first, we will find two fairly
simple solutions, and then, using these solutions, we will construct a
composite solution.

\textit{$1^\circ$. Solutions of exponential form in $x$}. Exact
generalized separable solutions of equation~\eqref{eqg:306} are sought
in the form
\begin{equation}
u(x,t)=\varphi(t)e^{\lambda x}+\psi(t),
\label{eqg:307}
\end{equation}
where functions $\varphi=\varphi(t)$ and $\psi=\psi(t)$ and the
constant $\lambda$ are to be determined in the subsequent analysis.
Substituting~\eqref{eqg:307} in~\eqref{eqg:306} and collecting similar
terms at exponents $e^{n\lambda x}$ ($n=0,\,1,\,2$), we get
\begin{equation*}
(b-2a\lambda^2)\varphi^2e^{2\lambda x}+
[\varphi'_t+(2b-a\lambda^2)\varphi\psi]e^{\lambda x}+\psi'_t+b\psi^2=0.
%\label{eqg:308}
\end{equation*}
Since this equality must be satisfied identically for any $x$, the
functional factors of $e^{n\lambda x}$ must be equated to zero. As
a result, we come to the differential-algebraic system
\begin{equation*}
\begin{aligned}
&b-2a\lambda^2=0,\\
&\varphi'_t+(2b-a\lambda^2)\varphi\psi=0,\\
&\psi'_t+b\psi^2=0,
\end{aligned}
%\label{eqg:309}
\end{equation*}
which allows two solutions
\begin{equation}
\lambda=\pm \BL(\frac{b}{2a}\BR)^{\!1/2},\quad
\varphi=\frac{C_1}{|t+C_2|^{3/2}},\quad
\psi=\frac 1{b(t+C_2)},
\label{eqg:310}
\end{equation}
where $C_1$ and $C_2$ \arbs.

\textit{$2^\circ$. Composite solution of exponential form in $x$}. From
relations~\eqref{eqg:307} and~\eqref{eqg:310} it follows that
equation~\eqref{eqg:306} has two solutions $u_{1,2}=\varphi
e^{\pm\lambda x}+\psi$. They differ in structure from each other only
by the sign of the exponent~$\lambda$.

This circumstance suggests trying to construct a more general
solution of equation~\eqref{eqg:306}, which includes both exponential
terms at once. In other words, we are looking for a composite solution
of the form
\begin{equation}
u(x,t)=\varphi_1(t)e^{-\lambda x}+
\varphi_2(t)e^{\lambda x}+\psi(t),\quad \lambda=\BL(\frac{b}{2a}\BR)^{\!1/2}.
\label{eqg:311}
\end{equation}
Substituting~\eqref{eqg:311} in~\eqref{eqg:306}, after elementary
transformations we have
\begin{equation*}
[(\varphi_1)'_t+\tfrac32b\varphi_1\psi]e^{-\lambda x}
+[(\varphi_2)'_t+\tfrac32b\varphi_2\psi]e^{\lambda x}+
\psi'_t+b(2\varphi_1\varphi_2+\psi^2)=0.
\end{equation*}
Equating the functional factors of $e^{n\lambda x}$ ($n=0,\,\pm
1$) to zero, we arrive at the first-order ODE system
\begin{equation}
\begin{aligned}
&(\varphi_1)'_t+\tfrac32b\varphi_1\psi=0,\\
&(\varphi_2)'_t+\tfrac32b\varphi_2\psi=0,\\
&\psi'_t+b(2\varphi_1\varphi_2+\psi^2)=0.
\end{aligned}
\label{eqg:312}
\end{equation}
Thus, it has been proved that equation~\eqref{eqg:306} admits the
solution of the form ~\eqref{eqg:311}.

By excluding $\psi$ from the first two equations in~\eqref{eqg:312}, we
obtain the equality\break
$(\varphi_1)'_t/\varphi_1=(\varphi_2)'_t/\varphi_2$. This implies that
$\varphi_1=A\varphi(t)$, $\varphi_2=B\varphi(t)$, where $A$ and $B$
\arbs. Therefore, the generalized separable solution~\eqref{eqg:311} is
reduced to the form
\begin{equation}
u(x,t)=\varphi(t)(Ae^{-\lambda x}+Be^{\lambda x})+\psi(t),\quad \lambda=
\BL(\frac{b}{2a}\BR)^{\!1/2},
\label{eqg:311a}
\end{equation}
where the functions $\varphi=\varphi(t)$ and $\psi=\psi(t)$ are
described by the nonlinear system of two ODEs:
\begin{equation}
\begin{aligned}
&\varphi'_t+\tfrac32b\varphi\psi=0,\\
&\psi'_t+b(2AB\varphi^2+\psi^2)=0.
\end{aligned}
\label{eqg:312a}
\end{equation}
By excluding $t$, this autonomous system is reduced to one ODE, which
is homogeneous and therefore can be integrated ~\cite{polzai2018}. Note
that the system of equations~\eqref{eqg:312a} for $AB>0$ admits two
simple solutions
\begin{equation*}
\varphi=\pm\frac 1{3b\sqrt{AB}\,(t+C)},\quad \psi=\frac{2}{3b(t+C)},
%\label{eqg:312b}
\end{equation*}
which define the solution~\eqref{eqg:311a} in the form of a product of
functions of different arguments.

\textit{$3^\circ$. Solution of trigonometric type in $x$}. When writing
formulas~\eqref{eqg:311} and~\eqref{eqg:311a} implicitly it was assumed
that $ab>0$. For $ab<0$ we have
\begin{equation*}
\lambda=i\beta,\quad \beta=\BL(-\frac{b}{2a}\BR)^{\!1/2},\quad i^2=-1.
\end{equation*}
In this case, in solution~\eqref{eqg:311a}  instead of exponential
functions, trigonometric functions appear, i.e. it can be represented
in the form
\begin{equation}
u(x,t)=\varphi(t)[A_1\cos(\beta x)+B_1\sin(\beta x)]+
\psi(t),\quad \beta=\BL(-\frac{b}{2a}\BR)^{\!1/2},
\label{eqg:311a*}
\end{equation}
where $A_1$ and $B_1$ \arbs. Substituting~\eqref{eqg:311a*} into
equation~\eqref{eqg:306} and performing calculations similar to those
in Item~$2^\circ$, we obtain the following nonlinear system of ODEs for
the functions $\varphi=\varphi(t)$ and $\psi=\psi(t)$:
\begin{equation}
\begin{aligned}
&\varphi'_t+\tfrac32b\varphi\psi=0,\\
&\psi'_t+b[\tfrac12(A_1^2+B_1^2)\varphi^2+\psi^2]=0.
\end{aligned}
\label{eqg:312a*}
\end{equation}
This system allows for two simple solutions
\begin{equation*}
\varphi=\pm\frac 2{3b\sqrt{A_1^2+B_1^2}\,(t+C)},\quad \psi=\frac{2}{3b(t+C)},
%\label{eqg:312b*}
\end{equation*}
which determine the solution~\eqref{eqg:311a*} in the form of a product
of functions of different arguments.
}

%\hypertarget{hlink-s:4}{}
\section{The Use of Complex-Valued Parameters for Constructing \\  Exact Solutions}\label{s:4}

%\hypertarget{hlink-ss:4.1}{}
\subsection{Linear Partial Differential Equations}\label{ss:4.1}

In the case of linear partial differential equations, the following
proposition can be used to construct more complex solutions from
simpler solutions.

\pproposition{\label{prd7}Let a linear homogeneous PDE with two
independent variables $x$ and $t$ have a one-parameter solution of the
form $u=\varphi(x,t,c)$, where $c$ is a parameter that is not included
in the original equation. Then the considered equation also has two
two-parameter solutions
\begin{equation}
u_1=\mathrm{Re}\, \varphi(x,t,a+ib),\quad \ \
u_2=\mathrm{Im}\, \varphi(x,t,a+ib),
\label{eq:blya77}
\end{equation}
where $a$ and $b$ are arbitrary real constants, $\mathrm{Re}\,z$ and
$\mathrm{Im}\,z$ are real and imaginary parts of complex number~$z$.}

\textbf{Proof.} The validity of the proposition follows from the
linearity of the equation and from the fact that the solution
$u=\varphi(x,t,c)$ is also a solution for $c=a+ib$.

Proposition~\ref{prd7} implies the validity of the following two
consequences:

\consequence{\label{prd71}Let a linear homogeneous PDE not depend
explicitly on the independent variable $t$ and have a solution
$u=\varphi(x,t)$. Then this equation also has two one-parameter
families of solutions
\begin{equation*}
u_1=\mathrm{Re}\, \varphi(x,t+ia),\quad \ \
u_2=\mathrm{Im}\, \varphi(x,t+ia),
\end{equation*}
where $a$ is an arbitrary real constant.}

\consequence{\label{prd72}Let a linear homogeneous PDE not depend
explicitly on the independent variable $x$ and have a solution
$u=\varphi(x,t)$. Then this equation also has two one-parameter
families of solutions
\begin{equation*}
u_1=\mathrm{Re}\, \varphi(x+ia,t),\quad \ \
u_2=\mathrm{Im}\, \varphi(x+ia,t),
\end{equation*}
where $a$ is an arbitrary real constant.}

\eexample{Consider the linear heat equation
\begin{equation}
u_{t}-u_{xx}=0.
\label{eq:002}
\end{equation}
It is easy to verify that this equation admits an exact solution of the
exponential form
\begin{equation*}
u=\exp(c^2 t+cx),
%\label{eq:003}
\end{equation*}
where $c$ is an arbitrary parameter.

Using Proposition~\ref{prd7}, we obtain two more complicated
two-parameter families exact solutions of equation~\eqref{eq:002}:
\begin{align*}
&u_1=\mathrm{Re}\,\exp(c^2 t+cx)|_{c=a+ib}=
\exp[(a^2-b^2)t+ax]\cos[b(2at+x)],\\
&u_2=\mathrm{Im}\,\exp(c^2 t+cx)|_{c=a+ib}=
\exp[(a^2-b^2)t+ax]\sin[b(2at+x)].
\end{align*}
}

\eexample{Consider the linear wave equation
\begin{equation}
u_{tt}-u_{xx}=0.
\label{f41}
\end{equation}
It is easy to verify that equation~\eqref{f41} admits translation
transformations for both independent variables and has the particular
solution
\begin{equation}
u=\frac{x}{x^2-t^2}.
\label{f42}
\end{equation}

Making a translation in solution~\eqref{f42} with an imaginary
parameter in $t$ and using Corollary~\ref{prd71}, we find
two more complicated one-parameter families of solutions to equation
\eqref{f41}:
\begin{equation*}
\begin{aligned}
&u_1=\mathrm{Re}\,\frac{x}{x^2-(t+ia)^2}=
\frac{x(x^2-t^2+a^2)}{(x^2-t^2+a^2)^2+4a^2t^2},\\
&u_2=\mathrm{Im}\,\frac{x}{x^2-(t+ia)^2}=\frac{2axt}{(x^2-t^2+a^2)^2+4a^2t^2}.
\end{aligned}
\end{equation*}

Making a translation in solution~\eqref{f42} with an imaginary
parameter in $x$ and using Corollary~\ref{prd72},
we find two other one-parameter families of solutions:
\begin{equation*}
\begin{aligned}
&u_3=\mathrm{Re}\,\frac{x+ia}{(x+ia)^2-t^2}=
\frac{x(x^2-t^2+a^2)}{(x^2-t^2-a^2)^2+4a^2x^2}\,,\\
&u_4=\mathrm{Im}\,\frac{x+ia}{(x+ia)^2-t^2}=
\frac{a(x^2+t^2+a^2)}{(x^2-t^2-a^2)^2+4a^2x^2}\,.
\end{aligned}
\end{equation*}
}

\eexample{Consider the linear heat equation
\begin{equation}
u_t=u_{xx}+\frac 1x{u_x},
\label{f411}
\end{equation}
which describes two-dimensional processes with axial symmetry, where
$x$ is the radial coordinate. It is easy to verify that
equation~\eqref{f411} admits a translation transformation with respect
to the variable $t$ and has the particular solution
\begin{equation}
u=\frac 1t\exp\Bl(-\frac{x^2}{4t}\Br).
\label{f421}
\end{equation}

Making a translation in solution~\eqref{f421} with an imaginary
parameter in the variable $t$ and using Corollary~\ref{prd71}, we find
two more complicated one-parameter families of solutions:
\begin{align*}
&\begin{aligned}
u_3&=\mathrm{Re}\,\frac{1}{t+ia}\exp{\BL(-\frac{x^2}{4(t+ia)}\BR)}={}\\
&=\frac{1}{t^2+a^2}\exp{\BL(-\frac{x^2 t}{4(t^2+a^2)}\BR)}
\BL(t\cos{\frac{ax^2}{4(t^2+a^2)}}+
a\sin{\frac{ax^2}{4(t^2+a^2)}}\BR),
\end{aligned}\\
&\begin{aligned}
u_4&=\mathrm{Im}\,\frac{1}{t+ia}\exp{\BL(-\frac{x^2}{4(t+ia)}\BR)}={}\\
&=\frac{1}{t^2+a^2}\exp{\BL(-\frac{x^2 t}{4(t^2+a^2)}\BR)}
\BL(a\cos{\frac{ax^2}{4(t^2+a^2)}}-t\sin{\frac{ax^2}{4(t^2+a^2)}}\BR).
\end{aligned}
\end{align*}
}

\eexample{Consider the linear wave equation with variable coefficients
\begin{equation}
u_{tt}-(xu_x)_x=0.
\label{f4}
\end{equation}
This equation admits a translation transformation with respect to $t$
and has the exact solution
\begin{equation}
u=\frac{Ct}{(4x-t^2)^{3/2}},
\label{f4**}
\end{equation}
where $C$ \arb.

By making a translation in solution~\eqref{f4**} an imaginary parameter
in $t$ and using the Corollary~\ref{prd71}, we can find
two more complicated one-parameter families of solutions by formulas
\begin{equation}
u_1=\mathrm{Re}\,\frac{C(t+ia)}{(4x-(t+ia)^2)^{3/2}}, \quad \
u_2=\mathrm{Im}\,\frac{C(t+ia)}{(4x-(t+ia)^2)^{3/2}}.
\label{eq:blya11}
\end{equation}
The final form of these solutions is not presented here, due to the
cumbersomeness of their recording. The solution $u_1$ was obtained
in~\cite{DobrT} and was used to describe the propagation of localized
disturbances in one-dimensional shallow water over an inclined bottom.
Note, that in~\cite{AksDobrDr}, another exact solution of the
equation~\eqref{f4} was obtained by integrating the parameter~$a$.
}

\pproposition{Let a linear homogeneous PDE have a one-parameter solution
of the form $u=\varphi(x,t,c)$, where $c$ is a real parameter that is
not included in the equation. Then, by $n$-fold differentiation or
integration of this solution, one can obtain other exact solutions of
the considered equation \cite{polman2007,polnaz2016}.}

\textbf{Corollary.} \textit{Exact solutions of linear PDEs, which do
not explicitly depend on the independent variable $t$, can be
constructed by differentiating or/and integrating with respect to
parameters $a$ and $b$ in solutions \eqref{eq:blya77} that are
obtained} by introducing the complex parameter $c=a+ib$.

\rremark{In~\cite{AksDobrDr} by integrating formulas~\eqref{eq:blya11}
with parameter $a$ it was obtained a new solution of
equation~\eqref{f4}.}

%\hypertarget{hlink-ss:4.2}{}
\subsection{Nonlinear Partial Differential Equations}\label{ss:4.2}

$1^\circ$. In some cases, it is possible to obtain another solution
from one solution, passing from real parameters to complex ones in such
a way that the transformed equation and the solution remain real. Let's
explain this with a few examples.

\eexample{Let us return again to equation~\eqref{eqg:306}. It is easy
to verify that its trigonometric solution~\eqref{eqg:311a*} and the
system of equations~\eqref{eqg:312a*} can be obtained from the solution
exponential form~\eqref{eqg:311a} and system of
equations~\eqref{eqg:312a}, if in the latter we formally set
\begin{equation}
\begin{aligned}
&e^{\lambda x}=e^{i\beta x}=\cos(\beta x)+i\sin(\beta x),\quad
e^{-\lambda x}=e^{-i\beta x}=\cos(\beta x)-i\sin(\beta x),\\
&A=\tfrac12(A_1+iB_1),\quad B=\tfrac12(A_1-iB),\quad A_1=A+B,\quad B_1=i(B-A).
\end{aligned}
\label{blya777}
\end{equation}
}

\eexample{Consider the equation
\begin{align}
u_t=au_{xx}+uf(u_x^2-bu^2),
\label{blya50}
\end{align}
where $f(w)$ \arbf.

It is easy to verify that equation~\eqref{blya50} has the simple
multiplicative separable solution (exponential in $x$):
\begin{align}
u=\psi(t)e^{\lambda x},
\label{blya51}
\end{align}
where the parameter $\lambda$ and the function $\psi(t)$ are to be
determined in the subsequent analysis. Substituting~\eqref{blya51}
in~\eqref{blya50}, we obtain two solutions of the form~\eqref{blya51},
where
\begin{align*}
\lambda=\pm \sqrt b,\quad \ \psi'_t=[ab+f(0)]\psi.
%\label{blya52}
\end{align*}
The presence of two solutions of the same type corresponding to $\pm
\lambda$ suggests trying them `combine' and look for a more general
composite solution of the form
\begin{align}
u=\psi(t)(Ae^{\lambda x}+Be^{-\lambda x}),
\label{blya53}
\end{align}
where $A$ and $B$ are some constants. Substituting~\eqref{blya53}
in Eq.~\eqref{blya50}, we obtain a solution of the
form~\eqref{blya53}, where $A$ and $B$ \arbs, and the function
$\psi=\psi(t)$ satisfies the nonlinear ODE:
\begin{align}
\psi'_t=ab\psi+\psi f(-4ABb\psi^2).
\label{blya54}
\end{align}

Substituting~\eqref{blya777} into~\eqref{blya53} and~\eqref{blya54}, we
arrive at a new solution containing already trigonometric functions
in~$x$,
\begin{align*}
u=\varphi(t)[A_1\cos(\beta x)+B_1\sin(\beta x)],\quad \ \beta=\sqrt{-b},
%\label{blya53*}
\end{align*}
where $A_1$ and $B_1$ \arbs, and the function $\psi=\psi(t)$ is
described by a nonlinear ODE:
\begin{align*}
\psi'_t=ab\psi+\psi f\bl(-(A_1^2+B_1^2)b\psi^2\br).
%\label{blya54*}
\end{align*}
}

\rremark{Exact solutions of the nonlinear hyperbolic equation
\begin{align*}
u_{tt}=au_{xx}+uf(u_x^2-bu^2)
\end{align*}
are constructed in the same way.}

$2^\circ$. Exact solutions of some nonlinear PDEs can be obtained using
the proposition below.

\pproposition{Let a nonlinear PDE have an exact solution involving
trigonometric functions of the form
\begin{equation}
u=F(x,t,A\cos(\beta x)+B\sin(\beta x),\beta^2),
\label{eqg:314*}
\end{equation}
where $A$, $B$, and $\beta$ are free real parameters that are not
included in the considered equation. Then this equation also has the
exact solution involving hyperbolic functions:
\begin{equation}
u=F(x,t,\bar A\cosh(\lambda x)+\bar B\sinh(\lambda x),-\lambda^2),
\label{eqg:314**}
\end{equation}
where $\bar A$, $\bar B$, $\lambda$ are free real parameters. The
converse is also true: if an equation has the exact
solution~\eqref{eqg:314**}, then it also has the exact
solution~\eqref{eqg:314*}.}

Solution~\eqref{eqg:314**} is obtained from~\eqref{eqg:314*} by
renaming the parameters $\beta=i\lambda$, $A=\bar A$, $B=-i\bar B$,
$i^2=-1$.

\eexample{Consider the fourth-order nonlinear equation
\begin{equation}
u_y(\Delta u)_x-u_x(\Delta u)_y=\nu\Delta\Delta u,\quad \
\Delta u=u_{xx}+u_{yy},
\label{eqg:314***}
\end{equation}
to which the stationary Navier--Stokes equations are reduced in the
planar case~\cite{loi1993}.

Equation~\eqref{eqg:314***} has the exact solution
\begin{equation*}
u(x,y)=[\bar A\sinh(\lambda x)+
\bar B\cosh(\lambda x)]e^{-\gamma y}+\frac\nu\gamma(\gamma^2+\lambda^2)x.
\end{equation*}
Therefore, this equation also has the exact solution
\begin{equation*}
u(x,y)=[A\sin(\beta x)+
B\cos(\beta x)]e^{-\gamma y}+\frac\nu\gamma(\gamma^2-\beta^2)x.
\end{equation*}
These solutions and other examples of this kind can be found
in~\cite{PolZ2}.
}

%\hypertarget{hlink-s:5}{}
\section{Using Solutions of Simpler Equations for Construct \\ Solutions
to Complex Equations}\label{s:5}

\textbf{Preliminary remarks.} It is often possible to use solutions of
simpler equations to construct exact solutions to complex differential
equations. In this section, we will illustrate the reasoning in such
cases for nonlinear PDEs (see Subsection~\ref{ss:5.1}), as well as for
more complex nonlinear partial functional differential equations (see
Subsections~\ref{ss:5.2}--\ref{ss:5.4}).

%\hypertarget{hlink-ss:5.1}{}
\subsection{Nonlinear Partial Differential Equations}\label{ss:5.1}

The following example shows how precise solutions of nonlinear
reaction-diffusion equations can be used to generate exact solutions
to wave type equations.

\eexample{Consider the reaction-diffusion equation with quadratic
nonlinearity
\begin{align}
u_t=a(uu_x)_x+bu,
\label{xeq:01}
\end{align}
which admits several simple exact solutions, which are given below
and expressed in elementary functions (see, for example,~\cite{PolZ2}).

$1^\circ$. The additive separable solution:
\begin{align}
u=-\frac{b}{6a}x^2+\psi(t),
\label{xeq:02}
\end{align}
where $\psi(t)=C\exp\bl(\tfrac 23bt\br)$ and $C$ \arb.

$2^\circ$. The multiplicative separable solution:
\begin{align}
u=\psi(t)x^2,
\label{xeq:03}
\end{align}
where $\psi(t)=-be^{bt}(6ae^{bt}+C)^{-1}$ and $C$ \arb.

$3^\circ$. The generalized separable solution:
\begin{align}
u=\psi_1(t)x^2+\psi_2(t),
\label{xeq:04}
\end{align}
where $\psi_1(t)=-be^{bt}(6ae^{bt}+C_1)^{-1}$ and $\psi_2(t)=
C_2e^{bt}(6ae^{bt}+C_1)^{-1/3}$, and $C_1$ and $C_2$ \arbs.

$4^\circ$. The generalized separable solution:
\begin{align}
u=\psi_1(t)x^2+\psi_2(t)\sqrt x,
\label{xeq:05}
\end{align}
where $\psi_1(t)=-be^{bt}(6ae^{bt}+C_1)^{-1}$ and $\psi_2(t)=
C_2e^{bt}(6ae^{bt}+C_1)^{-5/8}$, and $C_1$ and $C_2$ \arbs.

Let us now consider a nonlinear wave type equation with a quadratic
nonlinearity of the form
\begin{align}
u_{tt}=a(uu_x)_x+bu.
\label{xeq:06*}
\end{align}

Equations~\eqref{xeq:01} and \eqref{xeq:06*} differ only in the
order of the derivative with respect to $t$ in the left parts of the
equations. Since the right-hand sides of these equations involving
derivatives with respect to~$x$ are the same, it is natural to assume
that the power structure of solutions with respect to $x$ of both
equations will also be the same, and only the functional factors that
depend on $t$ will change for different powers of $x$.

In other words, we look for exact solutions of wave type
PDE~\eqref{xeq:06*} in the same form as solutions of reaction-diffusion
PDE~\eqref{xeq:01}. As a result, we get the following four exact
solutions of PDE \eqref{xeq:06*}:

$1^\circ$. The additive separable solution of the form \eqref{xeq:02},
where the function $\psi=\psi(t)$ is described by the ODE:
\begin{align*}
\psi''_{tt}=-\tfrac13b\psi+b\psi.
\end{align*}

$2^\circ$. The multiplicative separable solution of the form
\eqref{xeq:03}, where the function $\psi=\psi(t)$ is described by the
ODE:
\begin{align*}
\psi''_{tt}=6a\psi^2+b\psi.
\end{align*}

$3^\circ$. The generalized separable solution of the form \eqref{xeq:04},
where the functions $\psi_1=\psi_1(t)$ and $\psi_2=\psi_2(t)$ are
described by the ODEs:
\begin{align*}
&\psi_1''=6a\psi_1^2+b\psi_1,\\
&\psi_2''=2a\psi_1\psi_2+b\psi_2.
\end{align*}

$4^\circ$. The generalized separable solution of the form \eqref{xeq:05},
where the functions $\psi_1=\psi_1(t)$ and $\psi_2=\psi_2(t)$ are
described by the ODEs:
\begin{align*}
&\psi_1''=6a\psi_1^2+b\psi_1,\\
&\psi_2''=\tfrac{15}{4}a\psi_1\psi_2+b\psi_2.
\end{align*}
}

The considered example is a good illustration of a rather general fact,
which is a consequence of the results of~\cite{GSv} (see
also~\cite{PolZ2,PolZh}) and can be formulated as the following
proposition.

\pproposition{Let the evolution partial differential equation
\begin{align}
u_t=F[u],
\label{blya1}
\end{align}
where $F[u]\equiv F(u,u_x,\dots,u^{(n)}_x)$ is the nonlinear
differential operator in $x$, has a generalized separable solution of
the form
\begin{align}
u=\sum^m_{k=1}\psi_k(t)\varphi_k(x).
\label{blya2}
\end{align}
Then, the more complex partial differential equation
\begin{align}
L_1[u]=L_2[w],\quad \ \ w=F[u],
\label{blya3}
\end{align}
where $L_1$ and $L_2$ are any linear differential operators in $t$,
$$
L_1[u]=\sum^k_{i=0}a_i(t)u^{(i)}_t,\quad \ L_2[w]=
\sum^m_{j=0}b_j(t)w^{(j)}_t,
$$
also has the generalized separable solution of the form \eqref{blya2}
with the same functions $\varphi_k(x)$ (but with other functions
$\psi(t)$).}

\rremark{In the equations \eqref{blya1} and \eqref{blya3}, the nonlinear
operator $F$ can explicitly depend on the variables $x$ and $t$.}

Let us now give an example of constructing an exact solution that
cannot be obtained by using Proposition~7.

\eexample{Consider the $n$th-order nonlinear PDE:
\begin{equation}
u_{tt}=u F(u_x/u,u_{xx}/u,\dots,u_x^{(n)}/u),
\label{e:000**}
\end{equation}
which differs from~\eqref{e:000} only in the order of the derivative
with respect to $t$ on the left part of the equation.

We look for the solution of equation~\eqref{e:000**} in the same form
as the solution of equation~\eqref{e:000}.
Substituting~\eqref{d1010} in \eqref{e:000**}, for the function
$\varphi=\varphi(z)$ we obtain a nonlinear ODE:
\begin{equation*}
k^2\varphi+2kq\varphi_z'+q^2\varphi''_{zz}=
\varphi F(p\varphi_z'/\varphi,p^2\varphi_{zz}''/\varphi,
\dots,p^n\varphi_z^{(n)}/\varphi).
\end{equation*}
}

%\hypertarget{hlink-ss:5.2}{}
\subsection{Partial Differential Equations with Delay}\label{ss:5.2}

In biology, biophysics, biochemistry, chemistry, medicine, control
theory, climate model theory, ecology, economics, and many other areas
there are nonlinear systems, the rate of change of parameters of which
depends not only on the current state of the system at a given time,
but also on the state system at some previous time~\cite{wu1996}. The
differential equations that describe such processes, in addition to the
unknown function $u=u(x,t)$ also include the function $w=u(x,t-\tau)$,
where $t$ is time, $\tau>0$ is the constant delay. In some cases,
we consider situations where the delay depends on the time,
$\tau=\tau(t)$.

The presence of a delay significantly complicates the analysis of such
equations. Although nonlinear PDEs with constant delay allow solutions
of the traveling wave type $u=u(z)$, where $z=x+\lambda t$ (see, for
example, \cite{wu1996,mei2009,lv2015,polsor2015aml}), they do not allow
self-similar solutions of the form $u=t^\beta \varphi(x t^\lambda)$,
which often have simpler PDEs without delay.

More complex than traveling wave solutions, exact solutions of
nonlinear reaction-diffusion type equations with delay were obtained in
\cite{mel2008,pol2013,pol2014a,pol2014c,pol2014b,pol2014d,pol2014f,pol2019a,polsor2021,polsor2020**}.
Exact solutions of nonlinear Klein--Gordon type equations with delay
and related nonlinear hyperbolic equations are given in
\cite{pol2014g,polsorvyaz2015,long2016,polsor2020a,polsor2021,polsor2020**}.

Below, with specific examples, we will show how exact solutions of
nonlinear delay PDEs can be found by using solutions of simpler PDEs
without delay.

\eexample{Let us consider a nonlinear reaction-diffusion equation with
a constant delay,
\begin{align}
u_t=a(uu_x)_x+bw,\quad \ w=u(x,t-\tau).
\label{xeq:06}
\end{align}

Equation~\eqref{xeq:06} is more complicated than the ODE without delay
\eqref{xeq:01} and goes into it at $\tau=0$. The presence of
the delay in~\eqref{xeq:06} does not affect the nonlinear term
containing derivatives in $x$. Therefore, we can assume that the power
structure of solutions in~$x$ of both equations will be the same, and
only the functional factors that depend on $t$ will change.

In other words, we look for exact solutions PDE with
delay~\eqref{xeq:06} in the same form, as solutions simpler PDE without
delay~\eqref{xeq:01}. As a result, we get the following four exact
solutions of the nonlinear delay PDE~\eqref{xeq:06}:

$1^\circ$. The additive separable solution of the form \eqref{xeq:02},
where the function $\psi=\psi(t)$ is described by the linear delay ODE:
\begin{align*}
\psi'_t=-\tfrac13b\psi+b\bar\psi,\quad \ \bar\psi=\psi(t-\tau).
%\label{xeq:02*}
\end{align*}

$2^\circ$. The multiplicative separable solution of the form
\eqref{xeq:03}, where the function $\psi=\psi(t)$ is described by the
nonlinear delay ODE:
\begin{align*}
\psi'_t=6a\psi^2+b\bar\psi,\quad \ \bar\psi=\psi(t-\tau).
%\label{xeq:03*}
\end{align*}

$3^\circ$. The generalized separable solution of the form \eqref{xeq:04},
where the functions $\psi_1=\psi_1(t)$ and $\psi_2=\psi_2(t)$ are
described by the delay ODEs:
\begin{equation*}
\begin{aligned}
&\psi_1'=6a\psi_1^2+b\bar\psi_1,\quad \ \bar\psi_1=\psi_1(t-\tau),\\
&\psi_2'=2a\psi_1\psi_2+b\bar\psi_2,\quad \ \bar\psi_2=\psi_2(t-\tau).
\end{aligned}
%\label{xeq:04*}
\end{equation*}

$4^\circ$. The generalized separable solution of the form \eqref{xeq:05},
where the functions $\psi_1=\psi_1(t)$ and $\psi_2=\psi_2(t)$ are
described by the delay ODEs:
\begin{equation*}
\begin{aligned}
&\psi_1'=6a\psi_1^2+b\bar\psi_1,\quad \ \bar\psi_1=\psi_1(t-\tau),\\
&\psi_2'=\tfrac{15}{4}a\psi_1\psi_2+b\bar\psi_2,\quad \ \bar\psi_2=\psi_2(t-\tau).
\end{aligned}
%\label{xeq:05*}
\end{equation*}
}

\eexample{More complex than~\eqref{xeq:06}, nonlinear PDE with variable
delay
\begin{align*}
u_t=a(uu_x)_x+bw,\quad \ w=u(x,t-\tau(t)),
\end{align*}
where $\tau(t)$ \arbf, also admits four exact solutions of the form
\eqref{xeq:02}--\eqref{xeq:05}.
}

\eexample{The reaction-diffusion equation with logarithmic nonlinearity
\begin{align}
u_t=au_{xx}+u(b\ln u+c),
\label{eq:004cc}
\end{align}
admits the exact functional separable solution~\cite{GSv}:
\begin{align}
u(x,t)=\exp[\psi_2(t)x^2+\psi_1(t)x+\psi_0(t)],
\label{eq:004cd}
\end{align}
where the functions $\psi_n=\psi_n(t)$ are described by the nonlinear
system of ODEs:
\begin{align*}
&\psi_2^\prime=4a\psi_2^2+b\psi_2,\\
&\psi_1^\prime=4a\psi_1\psi_2+b\psi_1,\\
&\psi_0^\prime=a(\psi_1^2+2\psi_2)+b\psi_0+c.
\end{align*}

Let us now consider a more complex nonlinear reaction-diffusion
equation with a constant delay,
\begin{align}
u_t=au_{xx}+u(b\ln w+c),\quad \ w=u(x,t-\tau).
\label{eq:004cc*}
\end{align}

PDE with delay~\eqref{eq:004cc*} in the special case $\tau=0$ passes
into the simpler PDE without delay \eqref{eq:004cc}. For $\tau=0$ the
solution of delay PDE~\eqref{eq:004cc*}, as for
equation~\eqref{eq:004cc}, is sought in the form~\eqref{eq:004cd}. As a
result, for the functions $\psi_n=\psi_n(t)$, we obtain the
nonlinear system delay ODEs:
\begin{equation*}
\begin{aligned}
&\psi_2^\prime=4a\psi_2^2+b\bar\psi_2,\quad \bar\psi_2=\psi_2(t-\tau),\\
&\psi_1^\prime=4a\psi_1\psi_2+b\bar\psi_1,\quad \bar\psi_1=\psi_1(t-\tau),\\
&\psi_0^\prime=a(\psi_1^2+2\psi_2)+b\bar\psi_0+c,\quad \bar\psi_0=\psi_0(t-\tau).
\end{aligned}
%\label{eq:004cc**}
\end{equation*}
}

\eexample{More complex than~\eqref{eq:004cc*}, nonlinear PDE with
variable delay
\begin{align*}
u_{t}=au_{xx}+u(b\ln w+c),\quad \ w=u(x,t-\tau(t)),
\end{align*}
where $\tau(t)$ \arbf, also admits a solution with functional
separation of variables of the form~\eqref{eq:004cd}~\cite{pol2013}.
}

\rremark{In~\cite{polsor2021,polsor2020**}, methods for constructing
generalized traveling-wave solutions for nonlinear PDEs with delay by
using exact solutions more simple PDEs without delay were proposed.
This method is suitable for constructing exact solutions in both
explicit and implicit form.}

%\hypertarget{hlink-ss:5.3}{}
\subsection{Pantograph-Type Partial Differential Equations}\label{ss:5.3}

In this section, we will consider functional-differential equations
with partial derivatives of the pantograph type,
 which in addition to the unknown $u=u(x,t)$, also contain the same
functions with dilated or contracted arguments, $w=u(px,qt)$,
where $p$ and $q$ are scaling parameters (for equations with variable
delay we have $0<p<1$, $0<q<1$). Pantograph-type ODEs and PDEs are used
for mathematical modeling of various processes in
engineering~\cite{ock1971}, biology
\cite{hal1989,hal1991,der2012,zai2015,efe2018},
astrophysics~\cite{amb1944}, electrodynamics~\cite{deh2008}, the theory
of populations~\cite{aje1992}, number theory~\cite{mah1940}, stochastic
games~\cite{fer1972}, graph theory~\cite{rob1973}, risk and queuing
theories~\cite{gav1964}, the theory of neural networks~\cite{zha2013}.

Below, with specific examples, it is shown that the exact solutions of
nonlinear pantograph-type PDEs can be found by using simpler `ordinary'
PDEs, which do not contain the unknown function with dilated
or contracted arguments.

\eexample{Let us first consider a pantograph-type reaction-diffusion
equation with quadratic nonlinearity,
\begin{align}
u_t=a(uu_x)_x+bw,\quad \ w=u(px,qt).
\label{xeq:06***}
\end{align}

Equation~\eqref{xeq:06***} is more complicated than the ODE without
argument scaling~\eqref{xeq:01} and passes into it at $p=q=1$. The
presence in~\eqref{xeq:06***} of dilation in $w$ does not affect
the nonlinear term containing derivatives with respect to~$x$.
Therefore, we can assume that the power structure of solutions
in~$x$ of both equations will be the same, and only the functional
factors that depend on $t$ will change (and for the additive separate
solution, a factor in $x^2$ will change).

In other words, we look for exact solutions of pantograph-type PDE (93)
in the same form as solutions of `ordinary' simpler PDE~\eqref{xeq:01}.
As a result, we get the following four exact pantograph-type
solutions~\eqref{xeq:06***}:

$1^\circ$. The additive separable solution is
\begin{align*}
u=-\frac{bp^2}{6a}x^2+\psi(t);\quad \
\psi'_t=-\frac13bp^2\psi+b\bar\psi,\quad \ \bar\psi=\psi(qt).
\end{align*}

$2^\circ$. The multiplicative separable solution has the form
\eqref{xeq:03}, where the function $\psi=\psi(t)$ is described by the
pantograph-type ODE:
\begin{align*}
\psi'_t=6a\psi^2+bp^2\bar\psi,\quad \ \bar\psi=\psi(qt).
\end{align*}

$3^\circ$. The generalized separable solution has the form
\eqref{xeq:04}, where the functions $\psi_1=\psi_1(t)$ and
$\psi_2=\psi_2(t)$ are described by the pantograph-type ODEs:
\begin{align*}
&\psi_1'=6a\psi_1^2+bp^2\bar\psi_1,\quad \ \bar\psi_1=\psi_1(qt),\\
&\psi_2'=2a\psi_1\psi_2+b\bar\psi_2,\quad \ \bar\psi_2=\psi_2(qt).
\end{align*}

$4^\circ$. The generalized separable solution has the form
\eqref{xeq:05}, where the functions $\psi_1=\psi_1(t)$ and
$\psi_2=\psi_2(t)$ are described by the pantograph-type ODEs:
\begin{align*}
&\psi_1'=6a\psi_1^2+bp^2\bar\psi_1,\quad \ \bar\psi_1=\psi_1(qt),\\
&\psi_2'=\tfrac{15}{4}a\psi_1\psi_2+b\sqrt p\,\bar\psi_2,\quad \
\bar\psi_2=\psi_2(qt).
\end{align*}
}

\eexample{The reaction-diffusion equation with power-law nonlinearity
\begin{align}
u_t=au_{xx}+bu^k,
\label{eqp:001*}
\end{align}
for $k\not=1$ admits a self-similar solution~\cite{dor1982}:
\begin{align}
u(x,t)=t^{\ts\frac 1{1-k}}U(z), \quad \ z=xt^{-1/2},
\label{eqp:002}
\end{align}
where the function $U=U(z)$ is described by the nonlinear ODE:
\begin{align*}
\frac 1{1-k}U-\frac12zU'_z=aU''_{zz}+U^k,
\end{align*}

Let us now consider a much more complex nonlinear partial
functional-differential equation of the pantograph-type
\begin{align}
u_t=au_{xx}+bw^k,\quad \ w=u(px,qt),
\label{eqp:001}
\end{align}
where $p$ and $q$ are free parameters ($p>0$, $q>0$). Parameter values
$0<p<1$ and $0<q<1$ correspond to equations with proportional delay in
two arguments.

The functional-differential equation~\eqref{eqp:001} in the special
case $p=q=1$ passes into the `ordinary' partial differential
equation~\eqref{eqp:001*}. For $k\not=1$ the solution of the
pantograph-type PDE~\eqref{eqp:001}, as for equation~\eqref{eqp:001*},
is sought in the form~\eqref{eqp:002}. As a result, for the function
$U=U(z)$, we obtain a nonlinear ODE of the
pantograph-type~\cite{polsor2020p}:
\begin{align}
\frac 1{1-k}U-\frac12zU'_z=aU''_{zz}+bq^{\frac k{1-k}}W^k,\quad \ \
W=U(sz),\quad \ s=pq^{-1/2}.
\label{eqp:003}
\end{align}
}

\rremark{Equation~\eqref{eqp:001} with proportional delays for $0<p,q<1$
in the special case $p=q^{1/2}$ has an exact solution expressed in
terms of the solution of the ODE without delay~\eqref{eqp:003} with
$s=1$; for $p<q^{1/2}$, Eq.~\eqref{eqp:001} reduces to the delay ODE
with $s<1$; and for $p>q^{1/2}$, to the ODE with contracted argument
for $s>1$. Moreover, a solution of the ODE~\eqref{eqp:001} for $p,q>1$
for appropriate values of the parameters $p$ and $q$ can also be
expressed in terms of the solution of the ODE with delay ($s<1$),
without delay ($s=1$), and with contracted argument ($s>1$).}

\eexample{Let us now consider the reaction-diffusion equation with
exponential nonlinearity
\begin{align}
u_t=au_{xx}+be^{\lambda u}, \label{eqp:003x}
\end{align}
which for $\lambda\not=0$ admits the invariant solution~\cite{dor1982}:
\begin{align}
u(x,t)=U(z)-\frac 1\lambda \ln t,\quad \ z=xt^{-1/2}, \label{eqp:004}
\end{align}
where the function $U=U(z)$ is described by the nonlinear ODE:
\begin{align*}
-\frac 1\lambda-\frac 12zU'_z=aU''_{zz}+be^{\lambda U}.
\end{align*}

Let us now consider a much more complex nonlinear
functional-differential equation of the pantograph-type
\begin{align}
u_t=au_{xx}+be^{\lambda w},\quad \ w=u(px,qt),
\label{eqp:005}
\end{align}
where $p$ and $q$ are free parameters ($p>0$, $q>0$).

Partial functional-differential equation~\eqref{eqp:005} for
$p=q=1$ passes into the `ordinary' partial differential
equation~\eqref{eqp:003x}. For $\lambda\not=0$ the solution of the
pantograph-type equation~\eqref{eqp:005}, as for
equation~\eqref{eqp:003x}, is sought in the form~\eqref{eqp:004}. As a
result, for the function $U=U(z)$ we obtain a nonlinear ODE of the
pantograph-type \cite{polsor2020p}:
\begin{align*}
-\frac 1\lambda-\frac 12zU'_z=aU''_{zz}+\frac bq e^{\lambda W},\quad \ \
W=U(sz),\quad \ s=pq^{-1/2}.
\end{align*}
In the special case for $p=q^{1/2}$ this equation is a standard ODE
(without dilated or contracted arguments).
}

\eexample{It is easy to show that the nonlinear Klein--Gordon type
equation
\begin{align}
u_{tt}=au_{xx}+u(b\ln u+c)
\label{eqp:006}
\end{align}
allows the multiplicative separable solution
\begin{align}
u(x,t)=\varphi(x)\psi(t).
\label{eqp:007}
\end{align}

More complicated than~\eqref{eqp:006}, the nonlinear
pantograph-type PDE:
\begin{align}
u_{tt}=au_{xx}+u(b\ln w+c),\quad \ w=u(px,qt),
\label{eqp:2000}
\end{align}
also has the multiplicative separable solution~\eqref{eqp:007}, where
the functions $\varphi=\varphi(x)$ and $\psi=\psi(t)$ are described by
the nonlinear pantograph-type ODEs:
\begin{align*}
&a\varphi''_{xx}+\varphi(b\bar\varphi+c)=0,\quad \ \bar\varphi=\varphi(px);\\
&\psi''_{tt}=b\psi\ln\bar\psi,\quad \ \bar\psi=\psi(qt).
\end{align*}
}

\eexample{More complex than~\eqref{eqp:2000}, the partial
functional-differential equation
\begin{align}
u_{tt}=au_{xx}+u(b\ln w+c),\quad \ w=u(\xi(x),\eta(t)),
\label{xy*}
\end{align}
where $\xi(x)$ and $\eta(t)$ \arbfs, also allows a solution with the
separation of variables of the form~\eqref{eqp:007}.

In particular, for $\xi(x)=x-\tau_1$ and $\eta(t)=t-\tau_2$, where
$\tau_1$ and $\tau_2$ are some positive constants,
equation~\eqref{xy*} is a partial differential equation with two
constant delays.
}

%\hypertarget{hlink-ss:5.4}{}
%\subsection{The Principle of Analogy of Solutions}\label{ss:5.4}
\subsection{Approach for Constructing Exact Solutions of Functional \\
Partial Differential Equations}\label{ss:5.4}

Below, a rather general approach for constructing exact solutions of
functional partial differential equations of the pantograph-type is
formulated in the form following principle.

\textbf{The principle of analogy of solutions.}
%for pantograph-type PDEs.}
\textit{Structure of exact solutions to partial functional-differential
equations of the form
\begin{align}
F(u,w,u_x,u_t,u_{xx},u_{xt},u_{tt},\dots)=0,\quad \ w=u(px,qt)
\label{eq:xyzz}
\end{align}
often (but not always) is determined by the structure of solutions to
simpler partial differential equations:
\begin{align}
F(u,u,u_x,u_t,u_{xx},u_{xt},u_{tt},\dots)=0.
\label{eq:xyzzz}
\end{align}}

Equation~\eqref{eq:xyzzz} does not contain the unknown functions with
dilated or contracted arguments; it is obtained from~\eqref{eq:xyzz} by
formally replacing $w$ by $u$.

The solutions discussed in Examples 29--32 were constructed by using
the principle of analogy of solutions. Below are two more complex
examples.

\eexample{Consider the pantograph-type reaction-diffusion equation with
power-law nonlinearities
\begin{align}
u_t=au_{xx}+bu^mw^k,\quad \ w=u(px,qt),
\label{blya5}
\end{align}
that is more complex than \eqref{eqp:001}.

Following the principle of analogy of solutions, we set $w=u$ in
Eq.~\eqref{blya5}. As a result, we arrive at the equation
\begin{align*}
u_t=au_{xx}+bu^{m+k},
%\label{blya6}
\end{align*}
which, after renaming $m+k$ to $k$ coincides with Eq.~\eqref{eqp:001*}.
Taking into account that the solution of equation~\eqref{eqp:001*} is
determined by formula~\eqref{eqp:002}, the solution of
equation~\eqref{blya5} (by renaming $k$ to $m+k$ in
Eq.~\eqref{eqp:002}) are sought in the form \cite{polsor2020p}:
\begin{align*}
u(x,t)=t^{\ts\frac 1{1-m-k}}U(z), \quad \ z=xt^{-1/2},\quad \ k\not=1-m.
\end{align*}
As a result, for the function $U=U(z)$ we get the nonlinear
pantograph-type ODE:
\begin{align*}
&aU''_{zz}+\frac12zU'_z-\frac 1{1-m-k}U+bq^{\frac k{1-m-k}}U^mW^k=0,\\
&W=U(sz),\quad \ s=pq^{-1/2}.
\end{align*}
}

\eexample{Let us now consider the pantograph-type reaction-diffusion
equation with exponential nonlinearities
\begin{align}
u_t=au_{xx}+be^{\mu u+\lambda w},\quad \ w=u(px,qt),
\label{blya7}
\end{align}
that is more complex than \eqref{eqp:005}. Following the principle of
analogy of solutions, we set $w=u$ in Eq.~\eqref{blya7}. As a result,
we arrive at the equation
\begin{align*}
u_t=au_{xx}+be^{(\mu+\lambda)u},
%\label{blya8}
\end{align*}
which, after renaming $\mu+\lambda$ by $\lambda$ coincides with
Eq.~\eqref{eqp:003}. Taking into account that the solution of
equation~\eqref{eqp:005} is determined by formula~\eqref{eqp:004}, the
solution of equation~\eqref{blya7} (by renaming $\lambda$ to
$\mu+\lambda$ in Eq.~\eqref{eqp:004}) is sought in the form
\begin{align*}
u(x,t)=U(z)-\frac 1{\mu+\lambda} \ln t,\quad \ z=xt^{-1/2},\quad \
\mu\not=-\lambda.
\end{align*}
As a result, for the function $U=U(z)$ we obtain the nonlinear
pantograph-type ODE:
\begin{align*}
&aU''_{zz}+\frac 12zU'_z+\frac 1{\mu+\lambda}+
bq^{\ts-\frac{\lambda}{\mu+\lambda}} e^{\mu U+\lambda W}=0,\\
&W=U(sz),\quad \ s=pq^{-1/2}.
\end{align*}
}

\rremark{In~\cite{polsor2020p}, a number of exact solutions of nonlinear
pantograph PDEs of diffusion and wave types are obtained, which well
confirm the principle of analogy of solutions for pantograph-type PDEs.}
%(note that in \cite{polsor2020p} although the principle of analogy of
%solutions was used, but it was not formulated explicitly).}

%\hypertarget{hlink-s:6}{}
\section{Brief Conclusions}\label{s:6}

A number of simple, but quite effective, methods for constructing exact
solutions of nonlinear partial differential equations that require a
relatively small amount of intermediate calculations are described.
These methods are based on the following two main ideas: (i)~simple
exact solutions can serve as the basis for constructing more complex
solutions of the considered equations, (ii)~exact solutions of some
equations can serve as a basis for constructing solutions of more
complex equations. The effectiveness of the proposed methods is
illustrated by a large number of specific examples of constructing
exact solutions of nonlinear heat equations, reaction-diffusion
equations, wave type equations, hydrodynamics equations and some other
PDEs. In addition to exact solutions to partial differential equations,
some exact solutions to nonlinear delay PDEs and pantograph-type PDEs
are also described. The principle of analogy of solutions is
formulated, which allows us to constructively find exact solutions to
such partial functional-differential equations.

\textsl{Note that the simple methods and examples described in this article can be used in courses of lectures on equations of mathematical physics, methods of mathematical physics and partial differential equations for undergraduate and graduate students of universities.}

\renewcommand{\refname}{References}

\end{document}